\newif\ifFull
\Fullfalse
\documentclass[10pt]{article}
\usepackage{fullpage}
\usepackage{amsthm}
\usepackage{graphicx}
\usepackage{compress}

\title{Guard Placement For Wireless Localization}
\author{David Eppstein \hspace*{2em} Michael T. Goodrich \hspace*{2em} Nodari Sitchinava\\[5pt]
  Department of Computer Science\\
  University of California, Irvine \\
  { \tt \{eppstein, goodrich, nodari\}(at)ics.uci.edu}
}
\date{}

\newtheorem{theorem}{Theorem}
\newtheorem{lemma}{Lemma}

\newtheorem{conjecture}{Conjecture}

\begin{document}
\maketitle
\section*{Abstract}
Motivated by secure wireless networking, we consider the problem of
placing fixed localizers that enable mobile communication devices to
prove they belong to a secure region that is defined by the interior
of a polygon. Each localizer views an
infinite wedge of the plane, and a device can prove membership in the
secure region if it is inside the wedges for a set of localizers
whose common intersection contains no points outside the polygon.
This model leads
to a broad class of new art gallery type problems, for which we provide
upper and lower bounds.

\section{Introduction}
\emph{Localization}
is becoming an important topic in wireless mobile computing
(e.g., see~\cite{Cho-06}),
where we wish to determine with certainty the position of a wireless device
in a geometric environment.
Such localization problems are typically facilitated by 
\emph{locators}, which are wireless base stations placed at fixed locations
that aid the wireless devices to determine their positions.
In this paper, we are interested in such a
localization problem, where we are asked to deploy a collection of 
locators in what can be viewed as a two-dimensional space
so that a wireless device can prove that it belongs
to a given polygonal environment.
In this case, the locators are simple, fixed base stations that can broadcast
information in certain directions, such that devices outside of the
broadcast angle for a station cannot receive the transmissions
from that station.
Viewed geometrically, such a locator is a point-based \emph{guard} with a fixed
angle of view oriented in a fixed range of directions; hence, we typically
refer to such locators in this paper as ``angle guards'' or simply as
``guards'' if the context is clear.

From the standpoint of a mobile device,
which we model as a point $p$ in the plane,
an angle guard $g$ can be viewed as a Boolean
predicate, $B_g(p)$, which is true if $p$ is inside the angle associated with
$g$ and is false otherwise.
Moreover, we assume that if $B_g(p)$ is true, then the mobile device
associated with $p$ can produce a certificate for this fact.  For example,
the angle guard $g$ could periodically broadcast a secret key $K$ in its
transmission angle, so that only a wireless device in this angle would have
knowledge of this key (in which case a zero-knowledge
non-interactive proof-of-knowledge of $K$ would suffice 
as a certificate that $B_g(p)$ is true).

\begin{figure}[hbt!]
\ifFull\else\vspace*{-12pt}\fi
\begin{center}
\begin{tabular}{c@{\hspace*{3em}}c}
\includegraphics[scale=.7]{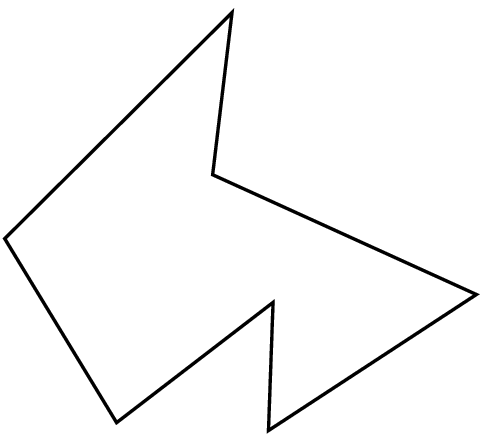}
&
\includegraphics[scale=.7]{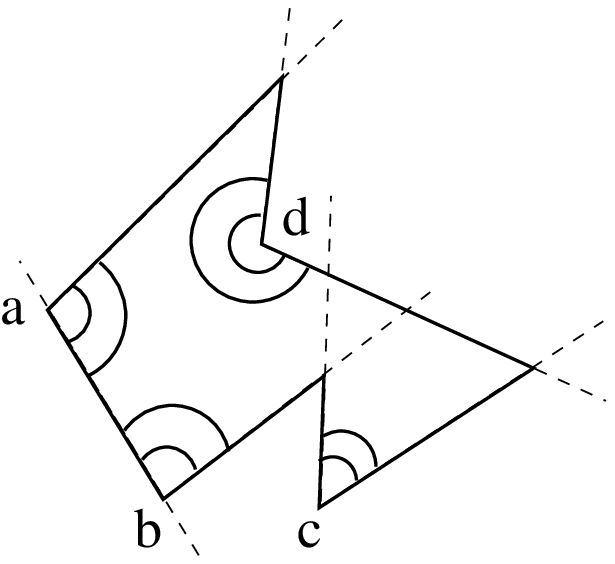} \\
(a) & (b) 
\end{tabular}
\ifFull\else\vspace*{-8pt}\fi
\caption{Illustrating the sculpture garden problem.
(a)~an example $7$-vertex polygon $P$; (b)~a set of $4$ 
angle guards that solve 
the sculpture garden problem for $P$.  The Boolean formula in this
case is $F=d(ab+c)$.}
\label{fig:example}
\end{center}
\ifFull\else\vspace*{-12pt}\fi
\end{figure}

Put strictly as a geometric problem, then, we are given a polygon $P$
and are asked to place angle guards in, around, and outside $P$ 
in such a way that we
can define a monotone\footnote{A Boolean formula is monotone if it
    contains only AND ($\cdot$) and OR ($+$) operators; hence, has
    no NOT operations.} Boolean formula, $F(p)$, built from the angle-guard
predicates, $B_g(p)$, so that $F(p)$ is true if and only if $p$ is inside $P$.
Moreover, we desire that the number of angle guards needed to define such a
formula be small, since there may be a non-trivial expense in deploying such
a collection of guards.
Thus, this problem can be viewed as a kind of art gallery
problem~\cite{ORo-87,o-v-97,u-agip-00}, 
where it is not sufficient that the guards merely
see all of the art gallery, but instead they must collectively define 
the geometry of the art gallery.
More specifically, this problem can be viewed as a ``sculpture garden''
problem, where the guards and the formula $F$ distinguish the space of the
sculpture garden from the surrounding land (without the use of walls or
fences).
(See Figure~\ref{fig:example}.)

Ideally, we would like the formula $F$ to be \emph{concise}, meaning that
$O(1)$ certificates always suffice to prove that $F(p)$ is true for
any point $p$ inside $P$.
The motivation for this desire is that we wish to prove that a point $p$ 
is inside $P$ using only angle-guard predicates,
and we would like that proof to be as short as possible.
For example, if $F$ were in disjunctive normal form (DNF), that is, $F$ was
a disjunction of conjunctive clauses, and each clause in $F$
contained a constant number of angle-guard predicates, 
then $F$ would be concise.

\subsection{Related Prior Work}
As mentioned above, localization is becoming an important topic in
wireless mobile computing, where a number of research teams are
interested in solutions that avoid the use of GPS, which has a number
of practical drawbacks.
For example, Bulusu {\it et al.}~\cite{bulusu00gpsless} study how RF
strength and angle can be used for sensor localization, and
Savvides {\it et al.}~\cite{savvides01dynamic} show how to improve
the consistency of such an approach by iterative algorithms.
Alternatively,
Howard {\it et al.}~\cite{howard02mobile} use a potential-field based
approach and
Chakrabarty {\it et al.}~\cite{chakrabarty02grid} use a grid-based
technique for deploying locators.
On the other hand,
He {\it et al.}~\cite{he03rangefree} use
a random deployment and use point-in-triangle tests to determine
location based on audible signals.

Of considerable relevance, of course, is prior work on using
directional antennas in wireless
communication networks.
For example, Ko and
Vaidya~\cite{ko00medium} discusses how to use base stations
with directional antennas (as in our angle guards) to improve network
protocols, but they assume that the mobile agents already know their
locations.
Bao and Garci-Luna-Aceves~\cite{bao02transmission}, on the other
hand, use directional antennas for adaptively discovering 
connection directions in an ad hoc network.
We are not familiar with any existing prior work, however, that uses
directional antennas for localization itself.
Nevertheless, using the results of our paper as a combinatorial
justification, a companion paper~\cite{Cho-06} addresses the
implementation issues of using locators with directional antennas
for mobile device localization.

Art gallery problems are a classic topic in Computational Geometry
and much has been written about them (e.g., see~\cite{ORo-87,o-v-97,u-agip-00}).
The starting point for this related research is a result of 
Chv{\'a}tal~\cite{c-ctpg-75} that
$\lfloor n/3\rfloor$ point guards are sufficient and sometimes
necessary to be able to fully see a simple polygon having $n$
vertices.
More related to the topic of this paper, ``prison yard''
problems~\cite{fk-pyp-94,ORo-87,o-v-97,u-agip-00}
seek a set of guards that can simultaneously
see both the interior and exterior of a simple polygon, in which case
$\lceil n/2\rceil$ guards are sufficient and sometimes 
necessary~\cite{fk-pyp-94}.
Relating to angle guards,
Estivill-Castro {\it et al.}~\cite{coux-ipwvl-95} show that vertex
angle guards (which they call ``floodlights'')
with angles of $180^\circ$ are sufficient to see any
simple polygon and there are polygons such 
that any fixed angle less than this will not.
Likewise, Steiger and Streinu~\cite{steiger98illumination}
and Bose {\it et al.}~\cite{bose97floodlight} study the complexity of
illuminating wedges with angle-restricted floodlights placed at a
fixed set of points.

Unfortunately,
solutions to art gallery or prison yard problems do not translate into
solutions to sculpture garden problems like the ones we study in this
paper, since we are interested in more than simply seeing the
inside and outside of a polygon---we wish to prove when a point is
inside a polygon using only the guards as witnesses.

Even more related to the topic of this paper is prior work on finding 
a constructive solid geometry (CSG)
representation of a simple polygon, since CSG representations can be
used to prove polygon containment.
Dobkin {\it et al.}~\cite{dghs-eafcr-93} describe a method for
constructing a formula $F$ that defines a simple polygon using 
primitives that are halfplanes defined by lines 
through polygonal edges, so that each halfplane is used exactly once.
Using our terminology, this is equivalent to a formula $F$ for
a set of $n$ angle guards, with each guard placed on an edge of the polygon
with a $180^\circ$ degree angle defined by the edge.
Such a formula would not, in general, be concise, however.
More recently, Walker and Snoeyink~\cite{ws-ppptu-99} study the
problem of using polygonal CSG representations, 
{\it a la} Dobkin {\it et al.}~\cite{dghs-eafcr-93},
for performing point-in-polygon tests.
They experimentally consider several interesting heuristics for improving the
efficiency of such tests, by ``flattening'' the CSG tree defined by
the formula, but they are not able to produce proofs
that are guaranteed to be concise in the sense of this paper.
Likewise, Goodrich~\cite{g-airsm-98} shows how
any CSG formula tree can be transformed into an equivalent
DAG of depth $O(\log n)$, but this again is not sufficient to
guarantee conciseness in the sense of this paper (in that we desire
constant-depth formulas).

Of course, one can always triangulate~\cite{Cha-DCG-91} any polygon,
$P$, and use two angle guards to define each of
the resulting $n+2(h-1)$ triangles, where $h$ is the 
number of holes in $P$.  This would give rise to a concise
formula $F$ for defining $P$, but it uses at least $2n+4(h-1)$ 
angle guards, which is much higher than we would like.
Thus, the challenge is to find ways of producing polygon-defining formulas
that use fewer than $n$ angle guards and are hopefully also concise.

\subsection{Our Results}
In this paper, we present a number of results
concerning the kinds and number of angle guards needed to define various
polygons (we use $n$ throughout to refer to the number of vertices of
a given polygon).
Specifically, we show the following:
\begin{enumerate}
\item
Define a \emph{natural} angle-guard vertex placement to be one where
we place each angle guard at a vertex of the polygon with
the angle of that vertex as the angle of the guard
(as in Figure~\ref{fig:example}).
We show there is a polygon $P$ such that a natural angle-guard
vertex placement cannot fully distinguish between points
on the inside and outside of $P$ (even if we place a guard at every
vertex of $P$).
This negative result implies that there are cases when 
we must use Steiner points or Steiner angles for sculpture garden
problems.
\item
We show that,
for any polygon $P$, there is a set of $n+2(h-1)$ angle guards
and an associated concise formula $F$ for solving the sculpture
garden problem for $P$, where $h$ is the number of 
holes in $P$ (so a simple polygon can be defined with $n-2$ guards).
\item
We give a class of simple polygons
that we conjecture require $n-2$ angle guards for any solution to the
sculpture garden problem.
\item
We observe that,
for any convex polygon $P$,
there is a natural angle-guard vertex placement such that
$\lceil n/2\rceil$ guards are sufficient to solve
the sculpture garden problem for $P$, and we show this bound is
optimal for any general-position polygon 
(for which no two edges belong to the same line).
\item
We show that $\lceil n/2\rceil + O(1)$ angle guards are sufficient to solve
the sculpture garden problem for pseudo-triangles.
\item
We show how any solution to the sculpture garden problem can be made
concise with a small blow-up in the number of guards.
\item
We give an example of a class of simple (non-general-position) 
polygons that have sculpture garden solutions using $O(\sqrt{n})$ guards, 
and we show this bound is optimal to within a constant factor.
\item
We show how to find a guard placement whose size is within a factor of $2$
of the optimal number for any particular polygon.
\item
We show that,
for any orthogonal polygon $P$ (which is probably the most likely
real-world application), there is a set of $\lceil 3(n-2)/4\rceil$ 
angle guards
and an associated concise formula $F$ for solving the sculpture
garden problem for $P$.
\end{enumerate}
Thus, we feel this paper begins an interesting new branch of work on
polygon guarding problems.

\section{Natural Angle-Guard Placements}
\begin{figure}[hbt!]
\begin{center}
\begin{tabular}{c@{\hspace*{4em}}c}
\includegraphics[width=2in]{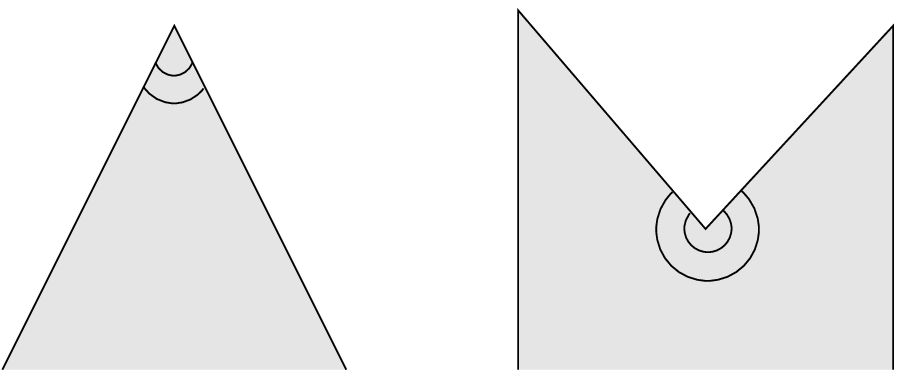}
&
\includegraphics[width=1.5in]{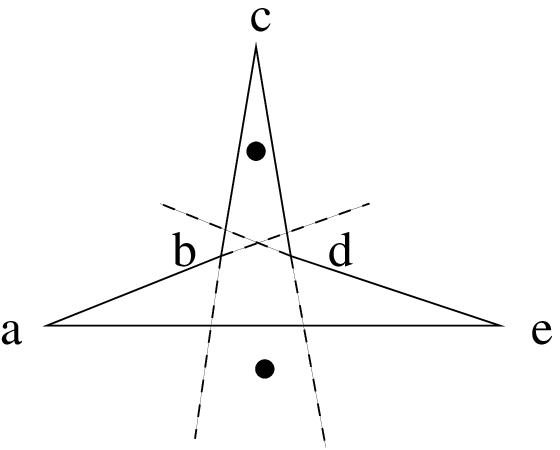} \\
(a) & (b)
\end{tabular}
\vspace*{-8pt}
\caption{Natural angle-guard placements. (a) examples of natural angle-guard 
vertex placements for convex and reflex angles of a polygon; (b) an
example polygon that cannot be defined using a natural angle-guard
placement, for the point, $p$, inside the polygon cannot be
distinguished from the point, $q$, outside the polygon.}
\label{fig:bad}
\end{center}
\vspace*{-4pt}
\end{figure}

As defined above,
a \emph{natural} angle-guard vertex placement is one where
we place each angle guard at a vertex of the polygon with
the angle of that vertex as the angle of the guard. 
(See Figure~\ref{fig:bad}a.)

A natural angle-guard placement has an obvious aesthetic appeal. Unfortunately,
the sculpture garden problem cannot be solved using natural
guards for some polygons. 

\begin{theorem}
There is a pentagon $P$ such that
it is impossible to solve the sculpture garden
for $P$ using a natural angle-guard vertex placement.
\end{theorem}
\begin{proof}
Let $P$ be the pentagon illustrated in Figure~\ref{fig:bad}b, and let
$p$ be the highlighted point inside of $P$ and let $q$ be the
highlighted point outside of $P$.
Then the natural guards cannot 
distinguish between the two points, $p$ and $q$.
For natural guards $a$ and $e$,
both points are outside the angles they cover, 
while, for guards $b$, $c$ and $d$, both points are
inside the angles which they cover.
That is, $B_x(p)=B_x(q)$, for $x=a,b,c,d,e$.
Therefore, any formula built using predicates $B_x$,
for $x=a,b,c,d,e$, will have identical values on $p$ and $q$.
Since $p$ and $q$ are on opposite sides of the boundary of $P$, this
implies that it is impossible to solve the sculpture garden problem
for $P$ using a natural angle-guard vertex placement.
\end{proof}

This theorem implies that some 
sculpture garden solutions
must use Steiner points or Steiner angles.
Nevertheless, for orthogonal polygons, natural guard placements suffice.

\begin{theorem}
Natural guards (one on each vertex) suffice to guard any orthogonal polygon.
\end{theorem}

\begin{proof}
Let $p$ be any point inside the polygon $P$.  We wish to show that the
intersection of the guarded regions for the natural guards containing
$p$, so for any $q$ outside $P$ we will find a guard separating $p$
from $q$.  To do so, let $R$ be the rectangle having $p$ and $q$ as
opposite corners.  The boundary $\partial P$ may cross $R$ many times,
but there is at least one component $B$ of $\partial P\cap R$ that
crosses $R$ and separates $p$ from $q$ in $R$, with $p$ on the interior
side of $B$; for instance, we may choose $B$ as the component of
$\partial P\cap R$ that is farthest from $p$ among the components
reachable from $p$ via paths in $P\cap R$.  Let $e$ be an edge of $B$
that crosses the boundary of $R$, and let $v$ be the endpoint of $e$
that is outside $R$; then the guard at $v$ separates $p$ from $q$.
\end{proof}

\section{An Upper Bound For Arbitrary Polygons}
In this section we show that the sculpture garden 
problem can be solved for any $h$-hole polygon 
with at most $n+2(h-1)$ guards and a concise formula. To prove this bound we
need to establish some preliminary results presented in the following lemmas.

\begin{lemma}
\label{lemma:tetragon}
The sculpture garden problem 
can be solved with two guards for any tetragon (quadrilateral).
\end{lemma}

\begin{figure}[hbt!]
\ifFull\else\vspace*{-4pt}\fi
\centering\includegraphics[width=2.5in]{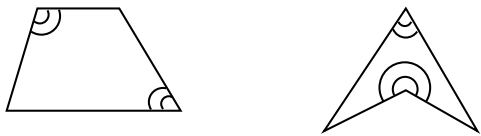}
\ifFull\else\vspace*{-8pt}\fi
\caption{Solutions for the sculpture garden problem for tetragons.}
\label{fig:tetragon}
\end{figure}

\begin{proof}
If the tetragon is convex, 
place the two natural angle-guards in any two opposite 
corners.
If the tetragon has a reflex vertex, place one natural 
angle-guard in the reflex vertex 
and the other in the opposite vertex (see Figure~\ref{fig:tetragon}).
The conjunction of the two angle guards defines the tetragons in each case.
\end{proof}

\begin{figure}[ht!]
\ifFull\else\vspace*{-12pt}\fi
\begin{center}
\begin{tabular}{c@{\hspace*{3em}}c@{\hspace*{3em}}c@{\hspace*{3em}}c}
\includegraphics[height=1in]{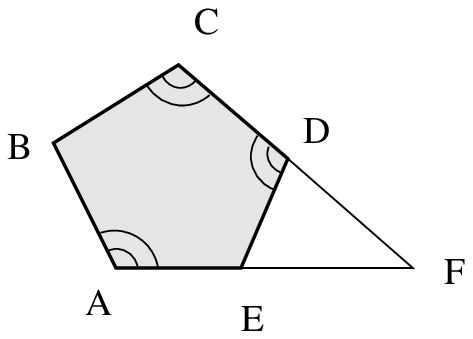} & 
\includegraphics[height=1in]{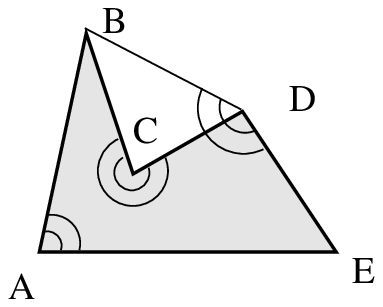} &
\includegraphics[height=1in]{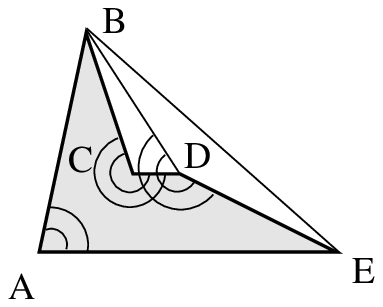} & 
\includegraphics[height=1in]{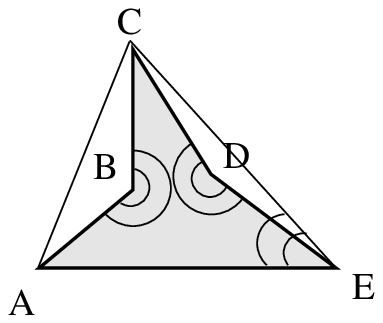} \\
(a) & (b) & (c) & (d)
\end{tabular}
\ifFull\else\vspace*{-4pt}\fi
\caption{Various pentagons $P = ABCDE$, as well as the corresponding convex 
  hulls $H$ and the containing tetragons $T$: (a) $H = ABCDE$, $T = ABCF$;
  (b) $T = H = ABDE$; (c) $H = ABE$, $T = ABDE$; (d) $H= ACE$, $T = ABCE$. 
  The corresponding solutions for the sculpture garden problem is the conjunction
  of all the guards placed.}
\label{fig:pentagon}
\end{center}
\ifFull\else\vspace*{-4pt}\fi
\end{figure}

\begin{lemma}
\label{lemma:pentagon}
The sculpture garden problem can be solved with three guards for any pentagon
$P$.
\end{lemma}
\begin{proof}
Consider a tetragon $T$ which fully contains 
the pentagon $P$ and shares at least $3$ consecutive edges of $P$. 
(We show later how to find $T$.) By Lemma~\ref{lemma:tetragon} we can 
solve the sculpture garden problem for the tetragon $T$ using exactly $2$ guards.

Now, since $T$ shares $3$ consecutive edges of $P$, it means that 
at least $4$ vertices of $P$ lie on $T$ or,
equivalently, there is at most $1$ vertex $v \in P$ that does not lie on $T$. 
That means that there are at most $2$ edges of $P$ which lie inside $T$ and 
which might not have been covered by guards. To complete the solution 
to the sculpture garden problem, place a natural angle guard at vertex $v$
(If there is no such vertex, i.e. only $1$ edge is not covered by the guards,
 it means that the pentagon $P$ is convex and we can place a natural angle 
 guard on either of the vertices incident on such an edge). 

The final solution to the sculpture garden problem on the pentagon will be 
the conjunction of all the guards placed for a total of $3$ guards.

To complete the proof we now describe how to find the tetragon 
$T$ which fully contains the pentagon $P$ and shares at least $3$ edges with it.
Consider the convex hull $H$ of $P$. 
\begin{itemize}
\item If $H$ consists of $5$ vertices (i.e. $P$ is
convex), pick any $4$ edges of $H$. $T$ is the tetragon which is constructed by the
intersection points of the lines on which those $4$ edges lie
(see Figure~\ref{fig:pentagon}(a)). Note that $3$ vertices of the tetragon
will be shared with the original pentagon. 
\item If $H$ consists of $4$ vertices, then $T$ is equal to $H$. 
\item If $H$ consists of $3$ vertices, then there are two cases to consider:
\begin{enumerate}
  \item {\em Two edges of the pentagon $P$ are also edges of $H$.} Note that
  the two edges have to be adjacent since $P$ is a pentagon.
  Let $ABCDE$ be the pentagon $P$ with vertices $A$, $B$, and $E$ comprising the
  vertices of the convex hull $H$ (see Figure~\ref{fig:pentagon}(c)). 
  Consider the edge $BE \in H$ which is not part of the pentagon $P$. Of
  the two pentagon vertices $C, D \not \in H$ at
  least one of them can be connected to both $B$ and $E$ without intersecting $P$.
  Without loss of generality let $D$ be such a vertex. 
  Since each one of $C$ and $D$ are adjacent to either vertex $B$ or $E$,
  one of the segments $DB$ or $DE$ is also an edge of the pentagon $P$ (in our
  example $DE \in P$). 
  Then the desired tetragon $T$ consists of the pentagon edges 
  $AB, AE$ and $DE$ as well as the segment
  $DB$. As desired, $T$ fully contains the pentagon $P$ (no edge of $T$ intersects
  $P$) and $T$ shares $3$ consecutive edges of $P$ ($AB, AE$ and $DE$ in our example).
  \item {\em Only one edge of the pentagon $P$ is also an edge of $H$.}
  Let $ABCDE$ be the pentagon $P$ with $ACE$ being the convex
  hull $H$ (see Figure~\ref{fig:pentagon}(d)). 
  Pick one of the two vertices $B, D \not \in H$. In our example we pick
  vertex $B$. 
  The desired tetragon $T$ consists of the pentagon edges 
  $AE$, $AB$ and $BC$ and the edge $CE$ of the convex hull $H$. 
  Note that the two vertices which are not on the convex hull
  ($B$ and $D$ in our example) will never be adjacent if the convex 
  hull shares only $1$ edge with the pentagon. Thus, both neighbors of each of
  those vertices are the vertices of the convex hull. Therefore, 
  the two rays originating from those vertices and shooting along the edges 
  of the pentagon ($BA$ and $BC$ in our example, since we picked $B$)
  don't intersect the pentagon $P$. Thus, the tetragon $ABCE$ fully contains
  the pentagon $P$ and shares $3$ consecutive edges ($AE$, $AB$ and $BC$) as desired. 
\end{enumerate}
\end{itemize}
\end{proof}

\begin{lemma}
\label{lemma:hexagon}
The sculpture garden problem can 
be solved with at most 4 guards for any hexagon.
\end{lemma}

\begin{proof}
Any hexagon whose dual graph of the triangulation is not a star graph or 
whose triangulation can be modified to have a non-star dual graph, can be
split into two tetragons each of which (by Lemma~\ref{lemma:tetragon}) can
be solved with two angle guards for a total of four. Thus, the only interesting
case is when a hexagon has a single triangulation and its dual graph is a 
star graph.

Let $H$ be such a hexagon and consider its triangulation. 
Since this is the only triangulation,
combining any pair of triangles produces non-convex tetragons. 
(If that wasn't the case,
we could combine two triangles into a convex tetragon and switch the diagonal to obtain
a different triangulation, which would violate the assumption of the uniqueness of the
triangulation.) 
Consider triangle $BDF$ which corresponds to the center vertex of the dual 
star graph. The lines on the boundary of the triangle $BDF$ partition the
plane into $6$ regions. For all pairs of adjacent triangles to construct a non-convex
tetragon it must be true that the vertices $A$, $C$ and $E$ lie in one of
the three shaded regions. Since at most $2$ of these vertices can lie in 
the same shaded region, there are two cases to consider: 

\begin{enumerate}
\item Each vertex $A$, $C$ and $E$ lie in its own region
(Figure~\ref{fig:star1}(a)). The vertices $B$, $D$ and $F$ are all reflex 
vertices and the rays originating at these vertices and shooting along
the edges of the polygon intersect each other only at the polygon 
vertices $A$, $C$ and $E$.
Thus, the conjunction of natural angle guards placed at the reflex vertices 
of the polygon (a total of $3$) will define the polygon 
(See Figure~\ref{fig:star1}(b)). 
\begin{figure}[hbt!]
\ifFull\else\vspace*{-12pt}\fi
\begin{center}
\begin{tabular}{c@{\hspace*{3em}}c}
\includegraphics[width=2in]{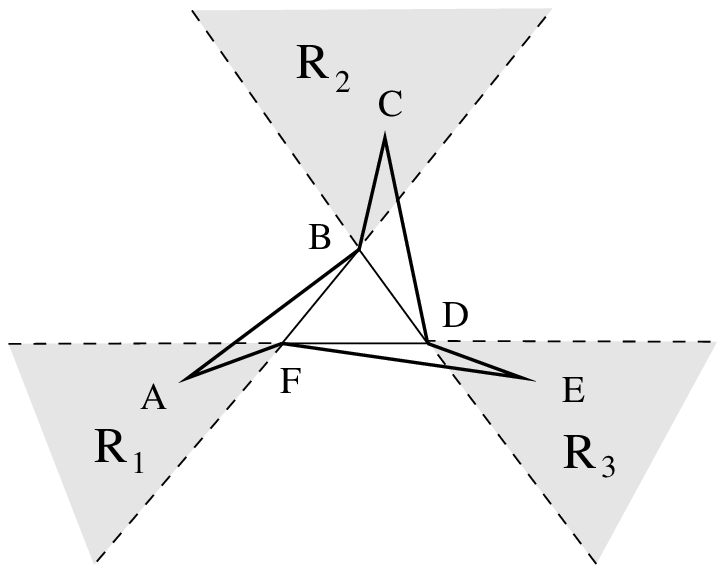} & \includegraphics[width=2in]{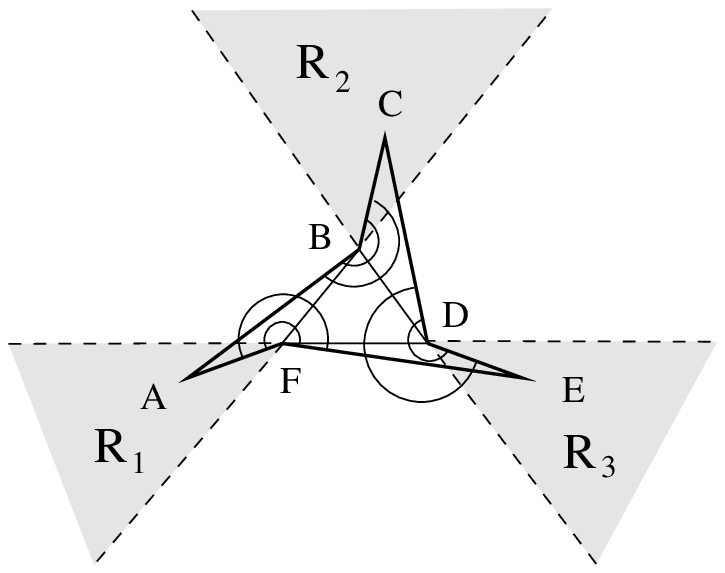} \\
(a) & (b)
\end{tabular}
\ifFull\else\vspace*{-4pt}\fi
\caption{An example of a hexagon with each of the vertices $A$, $C$ and $E$ 
  in their own region (a) and the corresponding solution for the sculpture 
    garden problem (b).}
\label{fig:star1}
\end{center}
\ifFull\else\vspace*{-4pt}\fi
\end{figure}
\item Two of the three vertices $A$, $C$, $E$ lie in the same region.
Without loss of generality let $A$ and $C$ lie in the same region $R_1$ and
vertex $E$ lie in region $R_3$ (Figure~\ref{fig:star2}(a)). 
Rays originating at vertices $A$ and $C$ and shooting along the edges of the polygon
all intersect the polygon edge $DF$. Consequently, they will all intersect edge
$EF$ and will never intersect edge $DE$ except at vertex $D$. Thus, the polygon
defined by the conjunction of two natural angle guards at vertices $A$ and $C$ 
and an edge guard\footnote{An {\em edge guard} is an angle guard with 
a $180^\circ$ angle defined by the edge on which it is placed.} 
on the edge $EF$ is fully contained inside the hexagon $H$. Moreover, the only part 
of the hexagon that is not covered by the above $3$ guards is a part of triangle 
$DEF$ near the vertex $E$. Since
we already have an edge guard at the edge $EF$ we can cover the whole triangle

$DEF$ by placing one additional angle guard at vertex $D$ whose wedge
is defined by the rays $DF$ and $DE$ (See Figure~\ref{fig:star2}(b)). 
Thus, a total of $4$ guards is required to guard this hexagon.
\end{enumerate}
\begin{figure}[hbt!]
\ifFull\else\vspace*{-12pt}\fi
\begin{center}
\begin{tabular}{c@{\hspace*{3em}}c}
\includegraphics[width=2in]{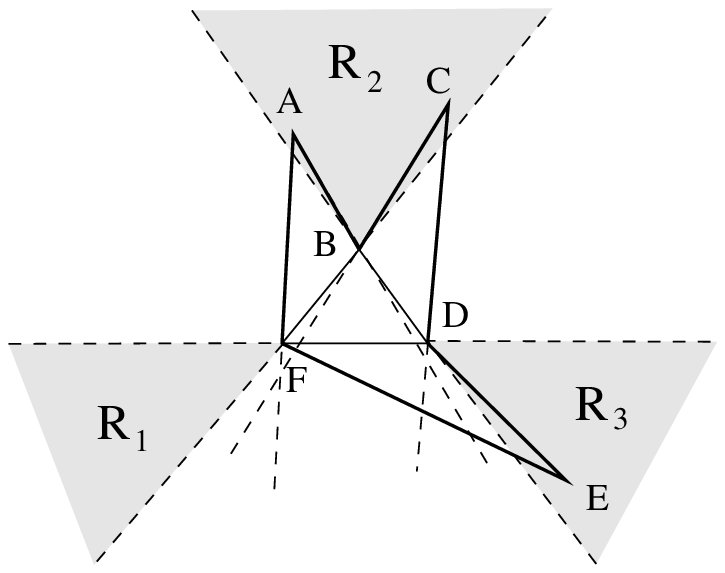} & \includegraphics[width=2in]{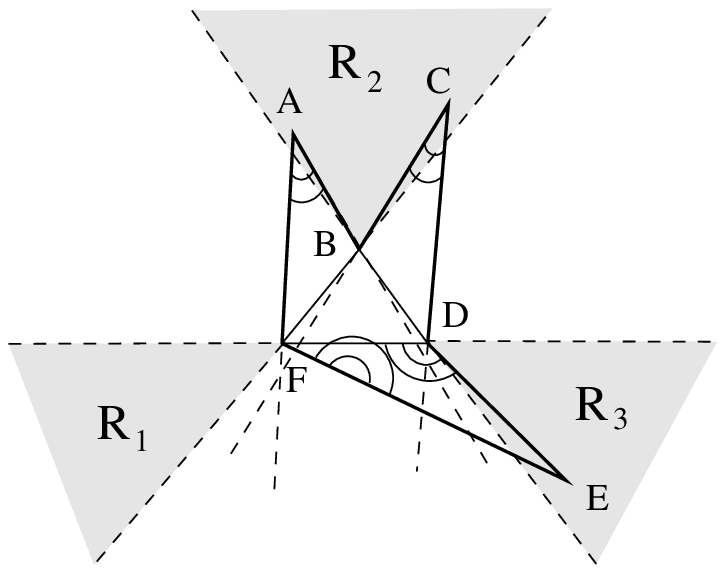} \\
(a) & (b)
\end{tabular}
\ifFull\else\vspace*{-4pt}\fi
\caption{An example of a hexagon with two vertices $A$ and $C$ in the same region 
  (a) and the corresponding solution for the sculpture garden problem (b).}
\label{fig:star2}
\end{center}
\ifFull\else\vspace*{-4pt}\fi
\end{figure}
Thus, we conclude that at most $4$ guards are required to guard any hexagon.
\end{proof}

\begin{lemma}
\label{lemma:partitioning}
Any polygon $P$ with more than three vertices can 
be partitioned into a collection of 
tetragons, pentagons and at most one 
hexagon whose dual triangulation tree is star-shaped.
\end{lemma}
\begin{proof}
Consider a dual spanning tree of a triangulation of the polygon $P$, which
is necessarily a degree-three tree.
If the tree is a two-, three- or
star-shaped four-node tree, we are done because the corresponding
polygon is a tetrahedron, a pentagon or a hexagon.

If there are more
than four nodes in the tree or the four-node tree is not star-shaped,
recursively trim the tree in the following way.
Pick a leaf $v$ such that $v$'s neighbor $u$ has one of the
following properties:
\begin{enumerate}
 \item $u$ has degree $2$ and $u$'s neighbor $w \neq v$ is not a leaf.
 \item $u$ has degree $3$ and {\em exactly one} of $u$'s other neighbors
       $w, z \neq v$ is also a leaf. Without loss of generality, let
       $w$ be an internal node, i.e. not a leaf.
\end{enumerate}
(Note, unless the tree is one of the base cases, a leaf $v$ with one of
the two properties always exists because the tree is a binary one.)

If $u$ has property $1$, then remove $v$ and $u$ from the tree and add the
tetragon, associated with the removed two nodes of the
tree into the collection. If $u$ has property $2$, then remove $u$, $v$, and
$z$ from the tree and add the pentagon associated with the removed
three nodes of the tree into the collection.

Continue the trimming until the tree is a two-, three- or
star-shaped four-node tree.  At each step we removed a tetragon or a
pentagon from the polygon $P$. Since we were removing only leaves with
their (common) neighbors at each step, the tree stays connected throughout the
trimming process. Therefore, the star-shaped four-node tree could have
emerged only at the end of the trimming process, i.e. there will be
only one hexagon.

There cannot be a single triangle left after the partitioning for the
following reason. A single triangle corresponds to a single node in
the dual tree. If there is any single node left after the trimming process
it would be $w$. However, in both properties $1$ and $2$ node $w$ is not a
leaf and, therefore, cannot be the only node left after the trimming.

Therefore, we can always partition the polygon into a collection of
tetragons, pentagons and at most one hexagon with a star-shaped dual
triangulation tree. 
\end{proof}

\begin{theorem}
$n+2(h-1)$ guards are sufficient to solve the sculpture garden problem 
with a concise formula with
the length of the proof certificate at most three for any polygon 
with $h$ holes.
\end{theorem}

\begin{proof}
Consider a triangulation of the polygon.  
Partition the polygon into
the collection of tetragons, pentagons and at most one hexagon as in
Lemma~\ref{lemma:partitioning}. Each tetragon will consist of two
triangles and by Lemma~\ref{lemma:tetragon} can be covered by two
guards. Each pentagon will consist of three triangles and by
Lemma~\ref{lemma:pentagon} can be covered by three guards. The hexagon
(if there is one) will consist of four triangles and by
Lemma~\ref{lemma:hexagon} can be covered by three guards. Thus, the
number of required guards will be no larger than the number of
triangles in the triangulation, which is $n+2(h-1)$.
The formula for the whole polygon will be the disjunction
of the formulas for each of the smaller polygons, which 
(by Lemmas~\ref{lemma:tetragon},~\ref{lemma:pentagon},~and~\ref{lemma:hexagon}) are 
conjunctions of length at 
most three. Thus, each proof certificate will be at most of length three. 
\end{proof}

\section{Lower Bounds}
In this section we discuss some lower bounds for sculpture garden problems.
We begin, however, with a conjecture.

\begin{conjecture}
There is a polygon that requires $n-2$ angle guards to solve the
sculpture garden problem.
\end{conjecture}

The following theorem establishes a lower bound on the number of guards for 
arbitrary polygons.

\begin{theorem}
\label{theorem:n/2}
At least $\lceil \frac n2 \rceil $ guards are required to solve the 
sculpture garden problem for any polygon with no two edges lying 
on the same line.
\end{theorem}
\begin{proof}
Assume less than $\lceil \frac n2 \rceil$ guards can guard a particular
polygon. Then there exists an edge $e$ which is not collinear with any of the
guards' boundary lines 
of the angle which they guard. This implies that there exists
a non-empty region $R$ which is fully located on one side (inside or outside) of each guards'
guarded region and such that edge $e$ splits $R$ into two subregions $R_1$ and
$R_2$. Without loss of generality assume $R_1$ is inside the polygon and $R_2$
is outside the polygon. Then no guard can distinguish whether a point is in
$R_1$ and $R_2$, i.e., no guard 
can distinguish between points inside and outside the
polygon. Thus, less than $\lceil \frac n2 \rceil$ guards cannot guard a
polygon. 
\end{proof}

Theorem~\ref{theorem:n/2} provides a general lower 
bound on the number of guards for an arbitrary general-position polygon, 
which is off by a factor of $2$ from the upper bound established above. 
For non-general-position polygons the following lower bound applies.

\begin{theorem}
\label{thm:sqrt-lower}
Any $n$-sided polygon requires $\Omega(\sqrt n)$ guards.
\end{theorem}

\begin{proof}
If a polygon $P$ is defined by $g$ angle guards, 
then $P$ can have at most $g(2g-1)$ polygon vertices, 
as each vertex occurs at the intersection of two of the $2g$ 
rays bounding guard regions.
\end{proof}

\section{Polygon Classes that Require Fewer than $n-2$ Guards}
In this section we consider classes of polygons for which the general
upper bound of $n-2$ guards for arbitrary polygons can be considerably
improved.  
%
\subsection{Convex Polygons}

We begin with an observation that,
for convex polygons, only $\lceil \frac n2 \rceil$ 
guards are required to solve the sculpture garden problem.

\begin{theorem}
\label{thm:convex}
$\lceil \frac n2 \rceil$ guards are always sufficient to solve the sculpture
garden problem for any convex polygon by placing the natural angle-guards in
every other vertex of the polygon.
\end{theorem}

\begin{proof}
Each natural angle-guard guards a region which fully contains the polygon. The
intersection of these regions is the convex hull of the polygon, which is the
polygon itself, since it is convex. Thus, the conjunction of 
the guards placed in every other corner of the convex polygon will define the
polygon itself.
\end{proof}

Together with the general lower bound on the number of guards, the above theorem
shows that $\lceil \frac n2 \rceil$ is a tight bound on the number of guards
required to solve the sculpture garden problem for convex polygons.
The formula is not concise, of course, 
but we show in
Section~\ref{sec:concise} how to make it concise with a small blow-up
in the number of guards.

\subsection{Pseudo-triangles}
A {\em pseudo-triangle} is a polygon with only 3 convex vertices and
the rest of the vertices being reflex. 
\begin{theorem}
Only $\lceil \frac n2 \rceil + O(1)$ guards are required to solve the 
sculpture garden problem for pseudo-triangles.
\end{theorem}

\begin{proof}
Insert a Steiner vertex $v$ anywhere in the kernel of the pseudo-triangle\footnote{A kernel
is a region of a star polygon (which pseudo-triangle is) from which one can
draw a line to any vertex without crossing the boundary of the polygon.}
and connect $v$ to each of the convex vertices of the polygon.
This partitions the polygon into three fans. 

Guard each fan separately by placing the natural angle-guards on every other 
reflex vertex of the fan, as well as at the newly created Steiner vertex $v$.
The formula for each fan will be the conjunction of the guard at $v$ and the
disjunction of all the guards on the reflex vertices.

The final formula will be the disjunction of the formulae of each separate
fan. See Figure~\ref{fig:pseudo-triangle} for an example.

\begin{figure}[hbt!]
\centering\includegraphics[width=2in]{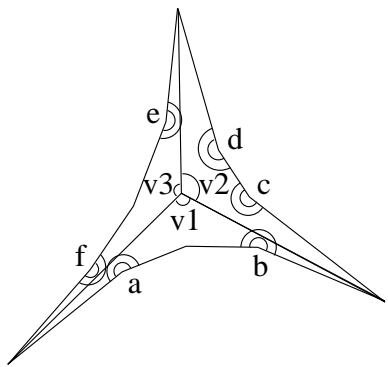}
\caption{An example of a pseudo-triangle and its guard coverage. The
corresponding formula is $F = v_1(a+b) + v_2(c+d) + v_3(e+f)$}
\label{fig:pseudo-triangle}
\end{figure}
The number of total guards to cover a pseudo-triangle is the number of guards
on the reflex vertices, which is $\lceil \frac n2 \rceil + O(1)$, plus three
more guards on the Steiner vertex $v$. Thus the total number of guards is
$\lceil \frac n2 \rceil + O(1)$.
\end{proof}

\subsection{Polygons with a Sublinear Number of Guards} 
We now present a class of polygons for which a square-root
number of guards is sufficient to solve the sculpture garden problem,
providing an upper bound within a constant factor of the lower bound
of Theorem~\ref{thm:sqrt-lower}.
\begin{theorem}
\label{thm:sqrt}
There exist $n$-sided simple polygons that can be guarded concisely 
by $O(\sqrt n)$ guards in a natural vertex placement.
\end{theorem}

\begin{figure}[hbt!]
\ifFull\else\vspace*{-4pt}\fi
\centering\includegraphics[width=1.5in]{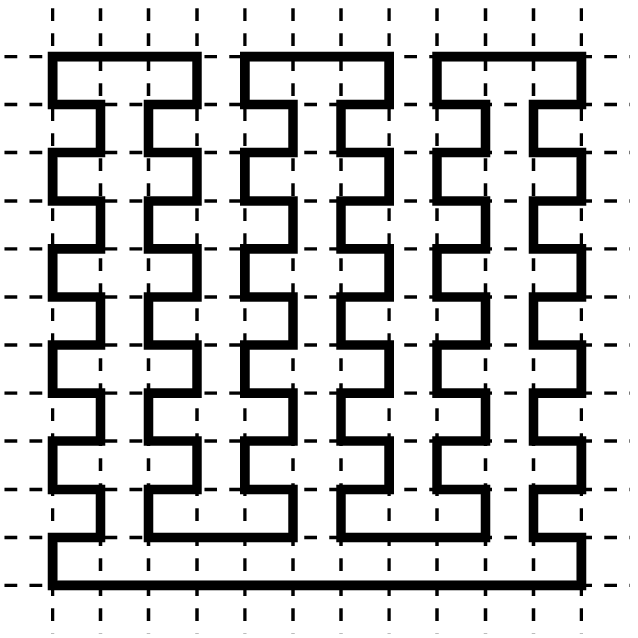}
\ifFull\else\vspace*{-8pt}\fi
\caption{An example polygon that can be defined with $O(\sqrt{n})$ angle
guards.}
\label{fig:zig}
\end{figure}

\begin{proof}
Form a line arrangement in the form of a grid with $4k$ horizontal
lines and $4k$ vertical lines, and let $P$ be a polygon with boundaries
that zigzag between pairs of vertical lines in the grid, as shown in
Figure~\ref{fig:zig}.  With such a construction we can form a vertex
of $P$ at every arrangement vertex except for some of the vertices on
the top and bottom horizontal lines of the arrangement, so $P$ has
$\Omega(k^2)$ vertices; by finding the next larger polygon of this form
and then simplifying it we can find for any $n$ a polygon with $n$
vertices, the edges of which belong to a grid with $k=O(\sqrt n)$.  We
place $16k$ guards, one on each side of each line of the
arrangement (using natural angle guards placed at vertices).
Using these guards, we can separately guard each rectangle
of the arrangement, and hence $P$, with four guards per point.
\end{proof}

A natural question raised by this example is whether it is always possible to 
find an angle-guard placement that minimizes the number of guards for
a particular polygon.  Although this problem may be NP-hard, 
we show in the next theorem that we can always achieve a $2$-approximation
for this problem.

\begin{theorem}
For any polygon $P$, we can find in linear time a collection of guards for 
$P$, using a number of guards that is within a factor of two of optimal.
\end{theorem}

\begin{proof}
For each halfplane for which a portion of the boundary of the halfplane
is used as one of the boundary edges of $P$, place an edge guard on the
line bounding the halfplane, and construct
the Peterson CSG formula~\cite{dghs-eafcr-93} for $P$.
In any collection of guards for $P$, each such
halfplane must be guarded by one of the two rays from one of the
guards, so the optimal number of guards is at least half the number of
guards used.
\end{proof}

It's tempting to look for an exact algorithm for guarding, using graph
matching to find halfplanes that can be paired up and covered by a
single guard, but 
  it may not be apparent which halfplanes can be paired or whether
such pairings are independent from each other. In addition, it is unclear whether
the optimal set of guards always has guard rays that lie along polygon edges, as would
be produced by such a matching algorithm.

\subsection{Orthogonal Polygons}

We now consider the case when the input is a polygon with axis-parallel
sides (i.e., an \emph{orthogonal polygon}).  In this case we can
considerably improve our $n-2$ bound on the number of guards needed for
general polygons.  Our construction may place guards interior to the
polygon, as well as on the boundary, and is based on the following
result known for its application to art gallery theory:

\begin{lemma}[O'Rourke et al.~\cite{Gyo-SJADM-86,ManWoo-TR-84,ORo-JGeom-83,ORo-87}]
\label{lem:orthogonal-partition}
Any simple orthogonal polygon with $n$ sides may be partitioned into
$\lceil(n-2)/4\rceil$ orthogonal polygons, each having at most six
sides.  Such a partition may be found in time bounded by that for
finding a horizontal visibility diagram of the polygon ($O(n)$ time, by
Chazelle's algorithm~\cite{Cha-DCG-91}).
\end{lemma}

We require a slight refinement of this lemma.  Note that, for an
$n$-sided orthogonal polygon, $n$ must always be even (there are
exactly as many horizontal edges as vertical) so we need only
distinguish between the case when $(n-2)/2$ is even and when it is
odd.

\begin{lemma}
\label{lem:orthogonal-partition-II}
Any simple orthogonal polygon with $n$ sides may be partitioned into
$\lceil(n-2)/4\rceil$ orthogonal polygons, each having at most six
sides.  If $(n-2)/2$ is odd, at least one of the polygons in the
partition has only four sides. Such a partition may be found in
time~$O(n)$.
\end{lemma}

\begin{proof}
If $(n-2)/2$ is even, we are done.  Otherwise, form the vertical
visibility decomposition of the polygon by extending a vertical line
segment across the polygon from each concave vertex to the nearest
opposite boundary.  This is a partition of the polygon into rectangles;
if we form a graph with one vertex per rectangle and one edge between
any two rectangles sharing a visibility edge, this graph is a tree.
Let $R$ be a rectangle forming a leaf in this tree, use $R$ as one node
in the partition, and apply Lemma~\ref{lem:orthogonal-partition} to the
remaining $(n-2)$-vertex polygon.
\end{proof}

\begin{figure}[hbt!]
\vspace*{-6pt}
\centering\includegraphics{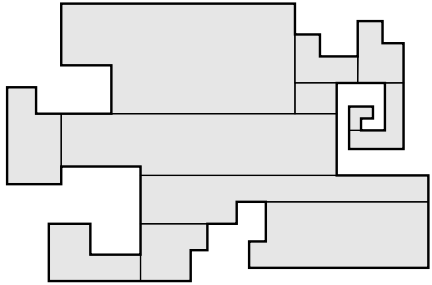}
\vspace*{-8pt}
\caption{Partition of an orthogonal polygon into rectangular and hexagonal pieces.}
\label{fig:Lpartition}
\end{figure}

Figure~\ref{fig:Lpartition} depicts the partition of an orthogonal
polygon into rectangular and hexagonal pieces according to these
lemmas.  Note that some of the vertices of the pieces may lie interior
to the polygon.

\begin{theorem}
\label{thm:orthogonal}
In any simple orthogonal polygon with $n$ sides, we may place at most 
$\lceil 3(n-2)/4\rceil$ right-angle guards, in such a way that any point within
the polygon has a concise proof of membership in the polygon, involving
only two guards.  This placement may be accomplished in time~$O(n)$.
\end{theorem}

\begin{proof}
We perform the partition of Lemma~\ref{lem:orthogonal-partition-II},
and guard each piece of the partition separately.  Each six-vertex
polygon must be an L-shaped union of two rectangles sharing a common
corner; it can be guarded by one guard at the common corner and one
guard at each of the two opposite rectangle corners, for a total of
three guards.  Each four-vertex polygon is a rectangle and may be
guarded with only two guards.  If $(n-2)/2$ is even, we obtain at most
$3(n-2)/4=\lceil 3(n-2)/4\rceil$ guards.  If $(n-2)/2$ is odd, we
obtain at most $2+3(n-4)/4=\lceil 3(n-2)/4\rceil$ guards.
\end{proof}

\section{Conciseness Trade-offs}
\label{sec:concise}
The formula we provided for convex polygons in
the proof of Theorem~\ref{thm:convex} is optimal as far as the number of
required guards goes. However, it is not concise; in fact, the
proof certificate is as long as the 
formula itself, i.e. $\lceil \frac n2 \rceil$.
This is far from the desired $O(1)$ bound for conciseness provided with other 
polygons in this paper. The following
theorem provides a trade-off between the number of required guards and the
conciseness of the formula.

\begin{theorem}
Let $P$ be a polygon taken from a class of polygons
that is closed under partitioning via diagonals
and such that $n$-vertex polygons of this class can be defined with $f(n)$
angle guards.
Then there is a concise solution to the 
sculpture garden problem for $P$ that uses $O(n f(c)/c)$ guards, where
$c$ is the maximum desired size of a proof a point is inside $P$.
\end{theorem} 

\begin{proof}
Triangulate $P$. If $P$ is not simple, then add diagonals so that the
dual to the triangulation is a tree $T$.
Perform a recursive centroid decomposition~\cite{g-psppt-95} of $T$,
stopping as soon as a subtree has size at most $c$.
Each cut of $T$ corresponds to our adding diagonals to $P$ and
this entire process introduces $O(n/c)$ subpolygons (of the same
class as $P$), each of size at most $c$.
Thus, each subpolygon can be defined with $f(c)$ angle guards, and we
can define a concise formula for $P$ that is the disjunction of the
formulas for the subpolygons.
\end{proof} 

For example, we can produce a concise guarding of a convex polygon
$P$ using $\lceil n/2\rceil(1+\epsilon)$ guards so that any point
can prove it is inside $P$ using $O(1/\epsilon)$ guards, for any
constant $\epsilon>0$.

\section{Conclusion and Open Problems}
In this paper, we introduced the sculpture garden problem for placing
angle guards in such a way as to define a polygon $P$ and prove when
points are inside $P$. We
presented a number of results concerning the kinds and number of guards
needed to define various polygons.  We provided the $n-2$ upper and
$\frac n2$ lower bounds for general polygons, as well as conjectured
the existence of some polygons which require as many as $n-2$ angle
guards.  
We also
provided several classes of polygons which require substantially fewer
guards than the general upper bound.
We feel this paper begins an interesting new branch of work on polygon 
guarding problems and hope that it will inspire future work in this direction.
In particular, we leave the following open problems:

\begin{enumerate}
\item
Is there a simple polygon
that requires $n-2$ angle guards to define it (our conjecture is ``yes'')?
\item
Our results also apply to the inverse sculpture garden problem,
so that mobile devices outside a polygon can prove they are outside.  
What are the best upper and lower bounds for a generalization of 
the sculpture garden problem so that devices inside or outside
the polygon can prove their respective locations?
\item
Establish tight upper and lower bounds for solving the sculpture
garden problem for orthogonal polygons.
\item
Is the problem of finding the minimum number of angle guards for a
particular polygon NP-hard?
\end{enumerate}

\subsection*{Acknowledgment}
We would like to thank Matthew Dickerson for helpful comments
regarding several of the proofs contained in this paper.

\small
\bibliographystyle{abbrv}
\bibliography{localize}
\end{document}